\documentclass[9pt,twocolumn,twoside]{osajnl}
\usepackage{amsmath}

\journal{jocn} 

\setboolean{shortarticle}{false}

\title{From Artificial Intelligence to Active Inference:\\The Key to True AI and 6G World Brain [Invited]}

\author[1,*]{Martin Maier}

\affil[1]{Optical Zeitgeist Laboratory, INRS, Montr\'eal, QC, Canada}

\affil[*]{Corresponding author: martin.maier@inrs.ca}

\begin{abstract}
In his opening OFC plenary talk back in 2021, Alibaba Group's Yiqun Cai notably added in the follow-up Q\&A that today's complex networks are more than computer science -- they grow, they are \emph{life}. This entails that future networks may be better viewed as techno-social systems that resemble biological superorganisms with \emph{brain-like} cognitive capabilities. Fast-forwarding, there is now growing awareness that we have to completely change our networks from being static into being a living entity that would act as an \emph{AI-powered network `brain'}, as recently stated by Bruno Zerbib, Chief Technology and Innovation Officer of France's Orange, at the Mobile World Congress (MWC) 2025. Even though AI was front and center at both MWC and OFC 2025 and has been widely studied in the context of optical networks, there are currently no publications on \emph{active inference} in optical (and less so mobile) networks available. Active inference is an ideal methodology for developing more advanced AI systems by biomimicking the way living intelligent systems work, while overcoming the limitations of today's AI related to training, learning, and explainability. Active inference is considered \emph{the key to true AI: Less artificial, more intelligent}. It is a biomimetic mathematical framework that is premised on first principles of statistical physics found in self-organizing/evolving complex adaptive systems, whether natural, artificial, or hybrid cyborganic ones. The goal of this paper is twofold. First, we aim at enabling optical network researchers to conceptualize new research lines for future optical networks with human-AI interaction capabilities by introducing them to the main mathematical concepts of the active inference framework. Second, we demonstrate how to move AI research beyond the human brain toward the \emph{6G world brain} by exploring the role of mycorrhizal networks, the largest living organism on planet Earth, in the AI vision and R\&D roadmap for the next decade and beyond laid out by Karl Friston, the father of active inference.
\end{abstract}

\setboolean{displaycopyright}{true}

\begin{document}

\maketitle

\section{Introduction}
\label{sec:intro}
This paper expands on my OFC 2025 invited paper titled ``From Artificial Intelligence to Active Inference: ``Natural Intelligence'' -- the Future of AI-Native 6G''~\cite{Maier25}, whose closing sentence states that, in the end, the acronym AI might actually not stand for artificial intelligence, but \emph{active inference} -- the key to true AI. While artificial intelligence was front and center at OFC 2025, it is worthwhile to mention that there was only one single paper inquiring into active inference. I have borrowed the closing statement above from a widely read WIRED article on Karl Friston, the world's most frequently cited neuroscientist as well as a friend and former colleague of cognitive psychologist and computer scientist Geoffrey Hinton -- the `Godfather of AI' and recipient of the 2024 Nobel Prize in Physics~\cite{Raviv18}. Active inference was pioneered by Karl Friston premised on the so-called \emph{Free-Energy Principle (FEP)}, a first principle of statistical physics. In stark contrast to the monstrous scaling energy demands of today's AI systems (e.g., large language models like ChatGPT), active inference facilitates the most energy efficient form of learning with no big data requirement necessary for training. Although active inference is still relatively young, it has a growing impact on various disciplines~\cite{PaPF22}. As we shall see, it is considered an ideal methodology for developing more advanced AI systems by biomimicking the way \emph{natural intelligence} works. Or, as the cover page of the aforementioned WIRED issue has titled: \emph{Less Artificial, More Intelligent.}\footnote{Speaking of \emph{Less Artificial, More Intelligent} -- to some, including the author of this article, it comes largely unsurprising that the WIRED article mentions that Karl Friston, a neuroscientist studying also mental disorder, doesn't have a mobile phone. There is now a growing awareness of the dangers of today's phone-based childhood in lieu of the analog play-based offline one, a topic in and of itself~\cite{Haidt24}.} 

The goal of this paper is twofold, which we shall call the \emph{Point Alpha} (short-term goal) and the \emph{Point Omega} (long-term goal) henceforth, respectively denoting the beginning and end of the active inference `Genesis' in optical networks in general and 6G networks in particular in this new \emph{Age of AI}~\cite{KiMS24}: 
\begin{description}
\item[Point Alpha -- Active Inference \& Optical Networks:] To my best knowledge, there are currently no publications on active inference in optical networks available, even though artificial intelligence and machine learning (AI/ML) have been widely studied in the context of optical networks, see, e.g.,~\cite{MRNM19} and~\cite{NPMM24} and references therein. Generally speaking, AI/ML excels at all sorts of pattern recognition related tasks in optical networks, ranging from quality of transmission (QoT) estimation of optical channels to anomaly detection in optical performance monitoring (OPM) systems. In recent years, AI/ML has become one of the key methods to automate a variety of optical network operations with the overarching goal of cost reductions and efficiency gains by removing the human out of the loop. At the downside, without the human in the loop, it will be difficult to fully capitalize on \emph{human-AI interaction}, a nascent research field of utmost importance not only for co-creating Ericsson's 6G vision of a cyber-physical world (see Fig.~\ref{fig1}) but also for scientific discovery in general, e.g., AI-powered drug discovery~\cite{Wang23}. Further, according to~\cite{NPMM24}, key AI research challenges remain open: ($i$) \emph{Training}: Lack of available training datasets from real-world network deployments, ($ii$) \emph{Learning}: Lack of lifelong (i.e., continual) learning, including AI degradation detection and model adaptation to progressive distribution shift, and ($iii$) \emph{Explainability}: Lack of trustworthy explainable AI (XAI) due to insufficient transparency of blackbox AI. 
\begin{figure}[t]
\centering
\includegraphics[width=\linewidth]{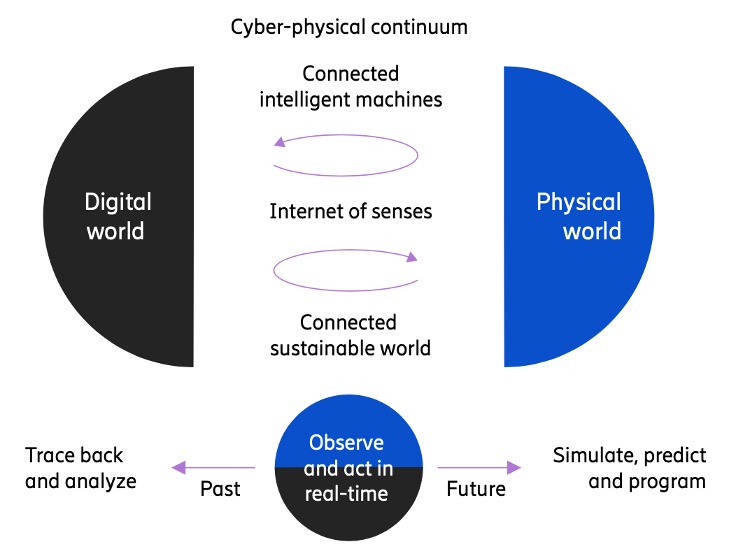}
\caption{Ericsson's 6G vision of co-creating a cyber-physical world: AI-native 6G is anticipated to move to a \emph{cyber-physical continuum} between the connected physical world of senses, actions, and experiences and its programmable digital representations~\cite{Ericsson24}. Note the similarity of the \emph{observe and act in real-time} cycle of AI-native 6G networks (bottom) to the \emph{action-perception cycle} of active inference (to be described shortly), though the term \emph{active inference} itself has not been mentioned in any Ericsson document up to date, indicating a certain degree of unawareness.}
\label{fig1}
\end{figure}

What's more, some liken the role of optical networks in today's high-speed networks to that of engines in high-speed cars. A car without a powerful engine will go nowhere. In fact, there is a wide consensus that 5G let alone 6G will go nowhere without suitable fiber backhaul infrastructures in place. The flip side of this analogy, however, is that optical network research might have reached its limits, similar to today's combustion engine whose efficiency has reached almost 100\%, leaving space for only incremental progress, unless new research lines are opened up, e.g., electrification of vehicles or, in the case of 6G, \emph{intelligentization} of networks -- that is, the ubiquitous deployment of AI to increasingly replace communication with computation. To see this, take the prime example of edge computing, which has been widely deployed for reducing network latency and/or predicting network traffic (in lieu of transmitting it). By introducing active inference as the key to true AI, this paper aims at enabling optical network researchers to tackle the aforementioned open key AI challenges of training, learning, and explainability, and arguably more importantly, providing them with a powerful framework to conceptualize new research lines for future optical networks with human-AI interaction capabilities premised on the so-called \emph{Markov blanket}, an important concept for modelling and designing interfaces between active inference agents interacting with each other in their surrounding environment (to be explained in more detail below).
\item[Point Omega -- Active Inference \& 6G World Brain:] In his opening OFC 2021 plenary talk, Alibaba Group's Yiqun Cai notably added in the follow-up Q\&A that today's complex networks are more than computer science -- they grow, they are \emph{life}. This entails that future networks such as 6G, Next G, and the 3D Metaverse -- the anticipated successor of today's mobile Internet -- may be better viewed as techno-social systems that resemble biological superorganisms with \emph{brain-like} cognitive capabilities~\cite{MEBR22}. This view has been recently echoed by Bruno Zerbib, Chief Technology and Innovation Officer of France's Orange, at the Mobile World Congress (MWC) 2025. In~\cite{LightReading25}, Zerbib argues that we have to completely change our networks from being static into being a \emph{living entity}. Zerbib goes on by saying that Orange is working on an AI-powered orchestration layer that would act as a \emph{network `brain'}, which is not expected to be fully formed at birth, but evolves over time. 

Since several years, we have been focusing on using AI to improve the intelligence and transmission performance and to reduce the complexity of communication systems, which is regarded as \emph{AI for communication network (AI4Net)}. AI4Net belongs to 5G and 5.5G. In the future, communication networks will be an integrated part of AI for performance optimization, in addition to data collection and transmission, which is known as \emph{communication network for AI (Net4AI)}. Net4AI is the mainstay for 6G and beyond~\cite{ToLi22}. In fact, it is anticipated that AI-native 6G will play a significant role in advancing Nikola Tesla's prophecy that \emph{``when wireless is perfectly applied, the whole Earth will be converted into a huge brain''}~\cite{SBGK19}.

Towards realizing this vision of a 6G world brain in the new Age of AI, we are going to explore which role Charles Darwin's \emph{``root-brain'' hypothesis} -- which has been forgotten in the literature until 2005 and fills a gap in the all-embracing \emph{Living Systems Theory} of James Grier Miller and is also in accord with Sir Jagadish Chandra Bose's \emph{Unity of Life} -- may play in advanced human-AI interaction of natural and artificial intelligences networked into a cyber-physical ecosystem premised on active inference.
\end{description}

The remainder of the paper is structured as follows. Section~\ref{sec:brain} briefly reviews the latest paradigm shifts in brain research, highlighting the major differences between outdated computer science and up-to-date neuroscience brain models. Section~\ref{sec:vision} then delves into the several limitations of today's AI and what it can become according to the AI vision and R\&D roadmap for 2030 and beyond recently outlined by Karl Friston \emph{et al.}, explaining what life is and how living systems in general learn, adapt, and self-evolve based on the FEP principle. Next, Section~\ref{sec:AIF} describes the active inference mathematical framework in more detail to provide the reader with a sufficient understanding of its main concepts, including but not limited to the aforementioned Markov blanket and generative model. In Section~\ref{sec:BeyondBrains}, we introduce the largest living system on planet Earth -- forests' underground mycorrhizal networks -- highlighted in Friston's AI vision and R\&D roadmap as a prime example of advancing future AI research beyond the human brain, which we subsequently investigate in light of the 6G world brain in Section~\ref{sec:6GWorldBrain}. Finally, Section~\ref{sec:conclusions} concludes the paper.

\section{The Brain: Computer Science vs. Neuroscience}
\label{sec:brain}
For decades, it has been the dominant metaphor to view the brain as a computer, drawing analogies from computer science such as that the brain processes input information and then makes a decision. This old ``outside-in'' strategy of the brain is now replaced with a new ``inside-out'' paradigm in modern neuroscience, where the \emph{brain does not process information: it creates it}~\cite{Buzs19}. In this new neuroscience dictum, the brain acts as a probabilistic prediction machine (rather than a deterministic processing machine) that predicts sensory inputs by means of action-oriented predictive processing. More specifically, the brain uses a \emph{generative model} to generate actions and predicts the sensory consequences in an ``inside-out'' manner of embodied action, while processing the actual sensory inputs as prediction errors. Or, put differently, the predictive brain creates a ``realm of possibilities'' (i.e., hallucinations) to perform a ``preplay of the future'' and then uses perception to control its hallucinations via action-based grounding (i.e., matching with reality). It's this \emph{action-perception loop}, briefly mentioned above in Fig.~\ref{fig1}, how living organisms continuously learn through active inference, which naturally lends itself to lifelong learning, a capability that is completely lacking in today's AI.

\begin{figure}[t]
\centering
\includegraphics[width=.75\linewidth]{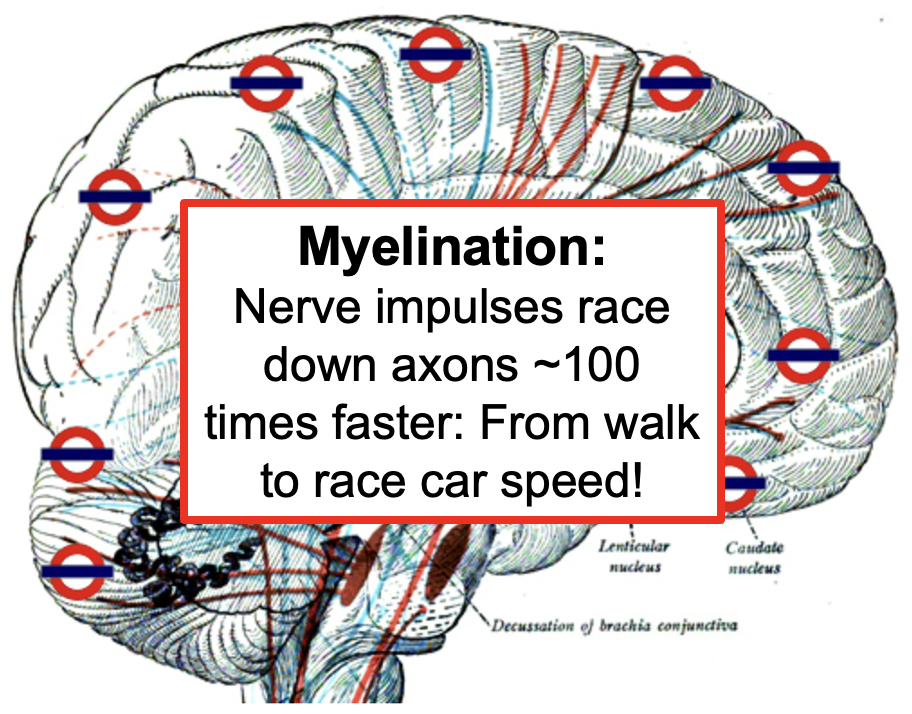}
\caption{White matter of the brain's subcortex: ``The subway of the brain.'' (Source: https://blogs.biomedcentral.com/on-biology)}
\label{fig2}
\end{figure}

DeepSeek, whose R1 model was launched recently in January 2025, seems to have realized how important the ``inside'' of the brain actually is, after the role of the brain's subcortex known as \emph{white matter} has been long underestimated. For decades neuroscientists showed little interest in white matter, which is underneath the gray matter -- the ``topsoil'' of the brain -- and its densely packed neurons. While gray matter makes up only 15\% of the human brain, white matter fills nearly half of it. Neuroscientists considered the white matter's millions of communication cables, each coated with a white substance called \emph{myelin}, for interconnecting neurons in one region of the brain with those in other regions little more than passive passageways, very much like the trunk lines that connect telephones in different parts of a country or, for the sake of a more modern example, network lanes in AI centers. Although scientists have long thought that white matter is passive tissue, new work shows that it \emph{actively} affects how the brain learns and dysfunctions (lack of myelin causes mental illnesses such as autism, schezophrenia, and multiple sclerosis). For instance, it was shown that nerve impulses race down myelin-coated axons on the order of 100 times faster; without myelination, the signal would leak and dissipate. That's the reason why white matter is sometimes referred to as ``the subway of the brain,'' as illustrated in Fig.~\ref{fig2}. Moreover, when learning a complex skill (e.g., playing piano), noticeable changes occur in white matter during interactions with an enriched environment. It was shown that a higher development of white matter structure correlates directly with a higher IQ. Further discoveries of white matter, buried deep underneath the gray matter ``topsoil,'' await unearthing by future experts~\cite{Fields08}.

Similar to white matter's role of coordinating how well brain regions work together, DeepSeek has put a particular focus on interconnecting AI GPUs through a high degree of parallelism with regard to model, data, pipeline, tensor, and experts. Due to the high degree of parallelism, DeepSeek didn't need to rely on Nvidia's advanced AI GPUs (recall that the US government has imposed sanctions on exporting Nvidia's advanced AI GPUs to China) for dramatically lowering the training costs of their R1 model down to \$6 million, as compared to OpenAI GPT-4's \$100 million. Importantly, DeepSeek's different approach has sent out shockwaves with the largest loss (\$600 billion) in US stock market history in their wake, making experts wonder whether there are better ways of advancing AI other than throwing ever increasing computation power at it, including its concomitant massive power consumption.

\section{Today's AI and What It Can Become: AI Vision and R\&D Roadmap for 2030 and Beyond}
\label{sec:vision}
\subsection{Free-Energy Principle}
In a panel at the 2024 World Economic Forum, two of the biggest names in AI -- computer scientist Yann LeCun and neuroscientist Karl Friston -- agreed that today's AI is profoundly lacking the key elements found in living intelligent systems in terms of perception, memory, reasoning, and generating actions~\cite{WEF24}. Where they differ was in how to get there. LeCun, a pioneer in Deep Learning (DL), argues that we don't know how to do learning or train systems any other way than DL, an engineering approach to AI invented decades ago. Conversely, Friston, the father of active inference, challenges the reliance on DL by advocating a new and better alternative, rooted in the \emph{Free-Energy Principle (FEP)}, which was briefly introduced above in Section~\ref{sec:intro}. The FEP is a unifying theory of brain function explaining how neurons and biological systems in general learn, adapt, and self-evolve in nature~\cite{Friston10}. It is essentially a mathematical formulation of how biological systems resist a natural tendency to disorder in the face of a constantly changing environment. 

\subsection{Negative Entropy: Paradoxically Something Very Positive}
\label{sec:entropy}
The FEP provides the answer to a fundamental question: How do self-organizing adaptive systems avoid surprising states for the sake of their survival? They can do this by minimizing their free energy, which can be shown to be an upper bound on the surprise of sensory states (see Section~\ref{sec:AIF} below for details). This means that the probability distribution of sensory states must have low \emph{entropy}, a measure of uncertainty denoting the average surprise (i.e., there is a high probability that a biological system will be in any of a small number of states, and a low probability that it will be in the remaining states). In other words, living systems are unique among natural systems because they manage somehow to resist the second law of thermodynamics, which states that entropy increases or remains constant. All other self-organizing systems, from snowflakes to solar systems, follow an inevitable and irreversible path to disorder. 

The FEP answer's Erwin Schr\"odinger's famous question \emph{What is Life?} by asserting that all living systems actively reduce disorder (i.e., entropy) of their sensory and physiological states by minimizing their free energy. In~\cite{Schroedinger68}, Schr\"odinger elaborated on the concept of entropy and the role it plays in the behavior of living organisms. By default, living organisms keep increasing their entropy until they approach maximum entropy, which corresponds to the dangerous state of death. However, living organisms are able to free themselves from the entropy they keep producing by drawing \emph{negative entropy} from their environment thru metabolism, i.e., exchange with environment. Paradoxically, negative entropy may be viewed as something very positive: Living organisms feed upon negative entropy, they can only stay alive by continually drawing negative entropy from their environment, i.e., extracting order from the environmnet -- or, as put by Schr\"odinger, by ``drinking orderliness from a suitable environment.'' According to~\cite{RaBF18}, the challenge lies in translating the theory of FEP into productive scientific practice and applying it to \emph{species without a brain, like fungi and flora} (to be further explored below in Section~\ref{sec:BeyondBrains}). 

\subsection{AI Vision and R\&D Roadmap: From Artificial to Natural Intelligence}
In stark contrast to the monstrous scaling energy demands of today's AI systems (e.g., large language models like ChatGPT), both LeCun and Friston agree on the fact that the FEP -- based on a principle of least action -- facilitates the most energy efficient form of learning with no big data requirement necessary for training. Unsurprisingly, active inference is considered an ideal methodology for developing more advanced AI systems by biomimicking the way \emph{natural intelligence} works in order to bridge the gap between artificial and natural intelligence. At the downside, however, LeCun remarks that ``we don't know how to do this today with AI systems. That's the problem we need to solve over the next few years.'' 

Recently, Friston \emph{et al.} laid out an AI vision and R\&D roadmap premised on active inference for the next decade and beyond~\cite{FRKT24}. Friston reiterated that the design of AI should be informed by, and aligned with, nature's time-tested methods and design principles demonstrated across systems of nested intelligences, capable of achieving multi-scale homeostasis (i.e., equilibrium) via nature's incredible coordination and communicative power. While acknowledging neuroscience as a key inspiration for AI research, Friston argues that we must move beyond brains and embrace the active and nested characteristics of natural intelligence in living organisms. Specifically, Friston mentions forests' underground \emph{mycorrhizal networks}, which create a mutually beneficial symbiosis between fungi and flora both briefly mentioned above, as a prime example to facilitate communication, learning, and memory in trees (to be described in more detail below in Section~\ref{sec:BeyondBrains}). He suggests that artificial general intelligence (AGI) and artificial super intelligence (ASI) will emerge from the interaction of intelligences networked into a \emph{cyber-physical ecosystem of artificial and natural intelligence}, in which humans are integral participants  -- what he calls ``shared intelligence.'' Key to such an ecosystem will be the dyadic interaction between artificial and natural intelligence, giving rise to \emph{human-AI interaction}, a nascent research field of utmost importance not only for creating Ericsson's 6G vision of a cyber-physical world but also for scientific discovery in general (see Section~\ref{sec:intro}).

\section{Active Inference: Less Artificial, More Intelligent}
\label{sec:AIF}
Active inference formalizes the predictive and enactive views of the brain and living organisms in general. Although active inference is still relatively young, it has a growing impact across various disciplines, e.g., human-AI interaction or more generally human-computer interaction (HCI), where both human and computer are modelled as mutually interacting active inference agents~\cite{MuWS24}. In this section, we describe the active inference mathematical framework in more detail to provide the reader with a sufficient understanding of its main concepts, including but not limited to the Markov blanket and generative model briefly mentioned in Section~\ref{sec:intro}. For additional information and more technical details, the interested reader is referred to~\cite{PePF24} for a recent overview of the history and future of active inference. For readers interested in using active inference in their own research, an outstanding step-by-step tutorial requiring a minimal background in mathematics and programming, including supplementary code for building active inference models from scratch, can be found in~\cite{SmFW22}. 

\subsection{Markov Blanket: Interface for Interaction \& Metamorphosis}
\begin{figure*}[ht]
\centering
\includegraphics[width=.8\linewidth]{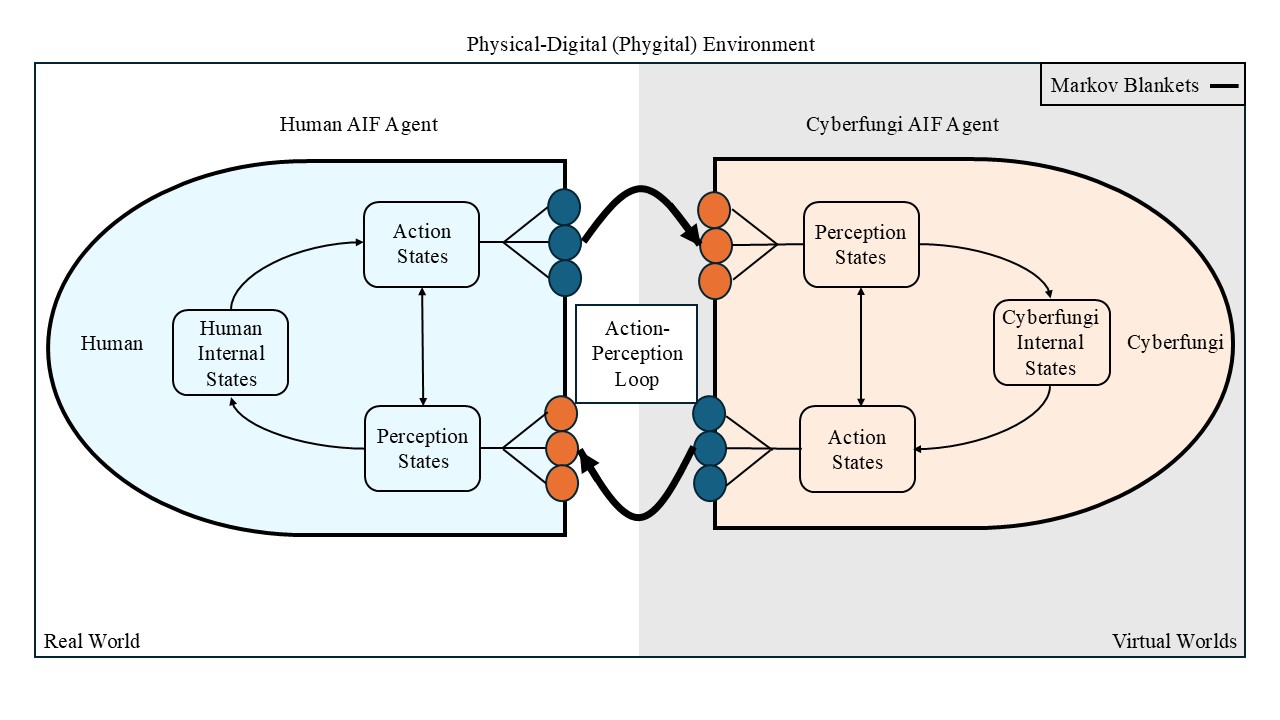}
\caption{Action-perception loop in a physical-digital (phygital) environment involving the dyadic interaction between human and digital active inference (AIF) agents, each with its own Markov blanket: A Markov blanket is a set of action and perception states that mediate all interactions between an AIF agent and its environment. Without Markov blanket, an AIF agent would dissolve into the environment, resulting in maximum entropy (i.e., death). Any living organism must enjoy some separation and autonomy from the environment by exchanging entropy with the environment to resist the second law of thermodynamics (i.e., negative entropy).}
\label{fig3}
\end{figure*}
The concept of Markov blanket forms the interface for interaction between an active inference agent and its surrounding environment by creating a coupling via action and perception states. It mathematically formulates the action-perception loop of an active inference agent via a Markov blanket $b$ given by
\begin{equation}
b=(u,y),
\label{equ1}
\end{equation}
where $u$ denotes the action states and $y$ denotes the perception states, respectively. The concept of Markov blanket was originally introduced in Judea Pearl's groundbreaking book \emph{Probabilistic Reasoning in Intelligent Systems: Networks of Plausible Inference}~\cite{Pearl88}. Similar to the Markov process underlying the widely studied Poisson traffic in networks research, the Markov blanket exhibits the Markov property, i.e., the memoryless property of a stochastic process where only the present state influences the probability distribution of future states. A Markov blanket generalizes the Markov property from the one-dimensional time domain of Poisson traffic to the three-dimensional space dimension of active inference agents interacting with their surrounding environment, including other active inference agents.

For illustration, Fig.~\ref{fig3} shows a physical-digital (phygital) environment (e.g., Ericsson's 6G vision of a co-created cyber-physical world in Fig.~\ref{fig1}), which hosts two active inference (AIF) agents, one human in the real world and another one in virtual worlds. The later one is assumed to be an advanced type of digital AIF agent referred to as \emph{cyberfungi}, which will be further investigated below in Section~\ref{sec:BeyondBrains}. Note that each AIF agent is characterized by its own Markov blanket, which separates the internal states of a given AIF agent from the external states of the surrounding environment, including other AIF agents. More complex AIF agents could also have multiple Markov blankets nested within one another (e.g., brains, organisms, communities), resulting in hierarchical Markov blankets with temporal and spatial depth. Any Markov blanketed system can be shown to engage in active inference~\cite{KPPF18}.

Markov blankets can grow and shrink depending on the mode of interaction, entailing a transition from Markov blanket states to processes. Active inference agents knit their own Markov blankets in ways that can change over time, leading to the concept of metamorphic agents. Where \emph{metamorphosis} occurs, the young life-stages do not look or behave in anything like the same way as the adult or mature life-stages. A familiar example of such metamorphic agents is the transformation of a caterpillar into a butterfly. Metamorphic insects account for at least 40\% of the world's total animal populations. Note that as an evolutionary strategy, metamorphosis works -- it is not a rare or exceptional solution to the problem of adaptive success thru a series of changes across the lifespan. In fact, metamorphosis has adaptive value because it allows younger and older living systems to share the same territory without consuming the same resources or being exposed to the same predators. Human beings are nature's experts at constantly re-configuring their own cognitive, bodily, and sensory boundaries of our \emph{socio-technological cocoon}, which enables us to re-invent a Markov blanket wherever new technologies interface with the old biological systems~\cite{Clark17}. Recent examples include virtual/augmented reality (V/AR) head-mounted devices for accessing the emerging 3D spatial Internet (also known as Metaverse) or novel types of human-AI interaction for the genesis of \emph{homo technicus}, a human species that may, in the new Age of AI briefly mentioned in Section~\ref{sec:intro}, live in symbiosis with machine technology (to be further explored below in Section~\ref{sec:6GWorldBrain}). 

It is important to note that fungi, briefly mentioned in our above discussion of negative entropy, are metamorphic agents as well. As we shall see in Section~\ref{sec:BeyondBrains}, fungi connect the roots of plants (flora) via underground pathways, though they may turn into mushrooms to entreat the visible more-than-fungal world above ground. Mushrooms are fungi's fruiting bodies, which let fungi embody themselves in the visible world. They are also the place where fungi produce spores to disperse themselves. Similar to nature's fungi, digital cyberfungi active inference agents should be able to entreat the visible more-than-virtual world (i.e., embodiment) and disperse themselves in the real world (i.e., agency), giving rise to \emph{embodied AI} and \emph{enactive AI}.

\subsection{Generative Model: Belief Update via Bayesian Inference}
\begin{figure*}[htbp]
\centering
\includegraphics[width=.8\linewidth]{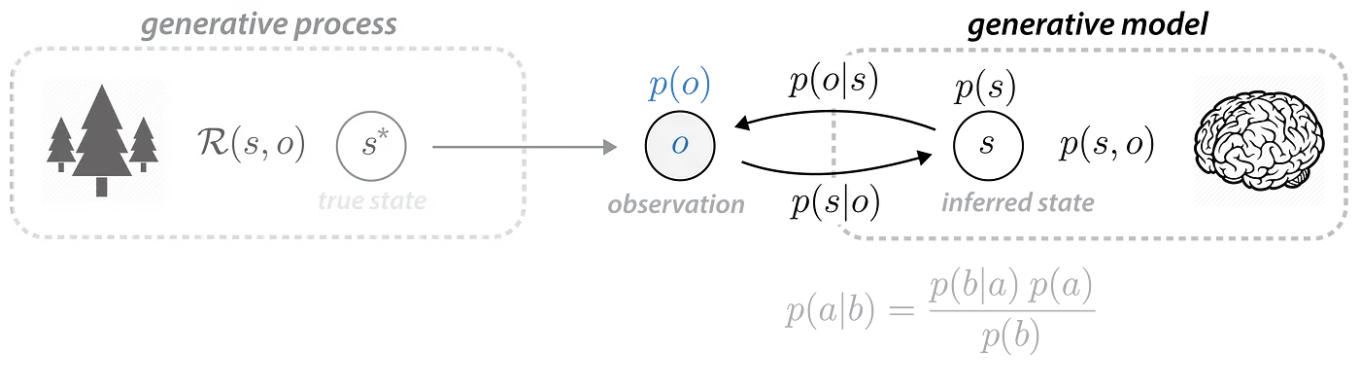}
\caption{At the heart of active inference lies a generative model: Getting the probabilistic generative model $p(s,o)$ right for updating an active inference agent's belief about the true state $s^*$ via Bayesian inference of state $s$ (i.e., cause $a$ in Bayes' theorem) from observation $o$ (i.e., consequence $b$ in Bayes' theorem) is the big challenge of active inference (Bayes' theorem is shown in light grey)~\cite{Solopchuk18}.}
\label{fig4}
\end{figure*}
Defining the Markov blanket for an active inference agent ensures that we know what is being inferred (the agent's external states outside its Markov blanket) and what is doing the inferring (the agent's generative model inside its Markov blanket). The generative model lies at the heart of active inference. Getting it right is the big challenge of active inference. The challenge is not to emulate the brain, but to find the generative model that describes the problem the brain is trying to solve. Once this is appropriately formalized in terms of a generative model, the solution to the problem emerges under active inference. Active inference can operate on different kinds of generative models. Therefore, the challenge is to specify the most appropriate form of the generative model for the problem at hand~\cite{PaPF22}. 

In active inference, it is important to distinguish between the generative model and the generative process that describes the dynamics of the world external to the active inference agent. As illustrated in Fig.~\ref{fig4}, the generative process corresponds to the process that determines the active inference agent's observation $o$. In many practical applications, the active inference agent's generative model is assumed to closely \emph{biomimic} the generative process that generates its observations~\cite{PaPF22}. As a result of its Markov blanket, an active inference agent is separated from external states, including the true state $s^*$ of the generative process that creates observation $o$. However, the agent can infer the hidden true state $s^*$ from its observation $o$ by using the well-known Bayes' theorem thru a process known as \emph{Bayesian inference} (see Fig.~\ref{fig4}). The inferred state $s$ after making observation $o$ is given by
\begin{equation}
p(s|o)=\frac{p(o|s)p(s)}{p(o)},
\label{equ2}
\end{equation}
where $p(s)$ denotes the \emph{prior belief} (encoded as probability distribution $p$) of different possible states $s$ \emph{before} making a new observation $o$. Whereas $p(s|o)$ denotes the \emph{posterior belief}, which encodes what an active inference agent's new belief (i.e., updated probability distribution $p$) optimally should be \emph{after} making a new observation $o$. Due to the fact that the internal states of any biological system are statistically insulated from the environment that generates sensory observations, an agent must engage in Bayesian inference about the hidden causes of its sensory states to behave optimally. In essence, Bayesian inference describes the optimal belief update of an active inference agent in light of a new observation (i.e., new sensory perception), not incorporating any action yet. Bayesian inference stands for the term `inference' in active inference.

It is important to keep in mind that Bayes' theorem is computationally intractable for anything but the simplest distributions. This is due to the fact that the number of all possible states and and their caused observations increases exponentially, rendering it intractable to compute $p(o|s)$ and $p(o)$ in Eq. (\ref{equ2}). As a consequence, approximation techniques are required that allow for approximate Bayesian inference. This is where \emph{free energy minimization} is crucial, as explained in the next subsection. 

Recall that the generative model lies at the heart of active inference. Typically, generative models use the concept of \emph{partially observable Markov decision process (POMDP)}, which offers a fairly expressive structure to model discrete state-space environments by means of parameters that can be expressed as tractable categorical distributions. A POMDP can be formally defined as a tuple of finite sets and matrices $(S, O, U, \bold{A}, \bold{B})$~\cite{PSCR24}: 
\begin{itemize}
\item $s\in S$: Set $S$ of states $s$ causing observations $o$.
\item $o\in O$: Set $O$ of observations $o$, where $o=s^*$ in fully observable settings (i.e., without Markov blanket) and $o=f(s)$ in partially observable settings (i.e., with Markov blanket); $f(s)$ correspond to perception states $y$ of the Markov blanket (see Eq. (\ref{equ1})), since generally perception differs from true state of reality.
\item $u\in U$: Set $U$ of actions $u$ of Markov blanket (see Eq. (\ref{equ1})).
\item $\bold{A}$: Matrix $\bold{A}$ encodes likelihood $p(o_\tau|s_\tau)$ at time $\tau$; an important thing to note here is that $\tau$ indexes the time points about which an agent has beliefs -- this is distinct from the below variable $t$, which denotes the time points at which a new observation is presented. Matrix has one column per state at $\tau$ and one row per possible observation at $\tau$.
\item $\bold{B}$: Matrix $\bold{B}$ encodes one-step transition dynamics $p(s_t|s_{t-1}, u_{t-1})$, i.e., probability that when action $u_{t-1}$ is taken while being in state $s_{t-1}$ at time $t-1$ results in $s_t$ at time $t$. It has one column per state at $t-1$ and one row per state at $t$.
\end{itemize}
Now, instead of passively inferring what caused observations without exerting any actions on the environment, active inference agents actively infer by using the available actions in $U$ of their POMDP-based generative model. Note that these actions stand for the term `active' in active inference. Also note that apart from the aforementioned POMDP matrices $\bold{A}$ and $\bold{B}$, a generative model may comprise additional matrices, including matrix $\bold{C}$ for preferred outcomes, $\bold{D}$ for prior beliefs about each possible state at initial time point $\tau=1$, and $\bold{E}$ for prior beliefs about habits.

\begin{figure}[t]
\centering
\includegraphics[width=\linewidth]{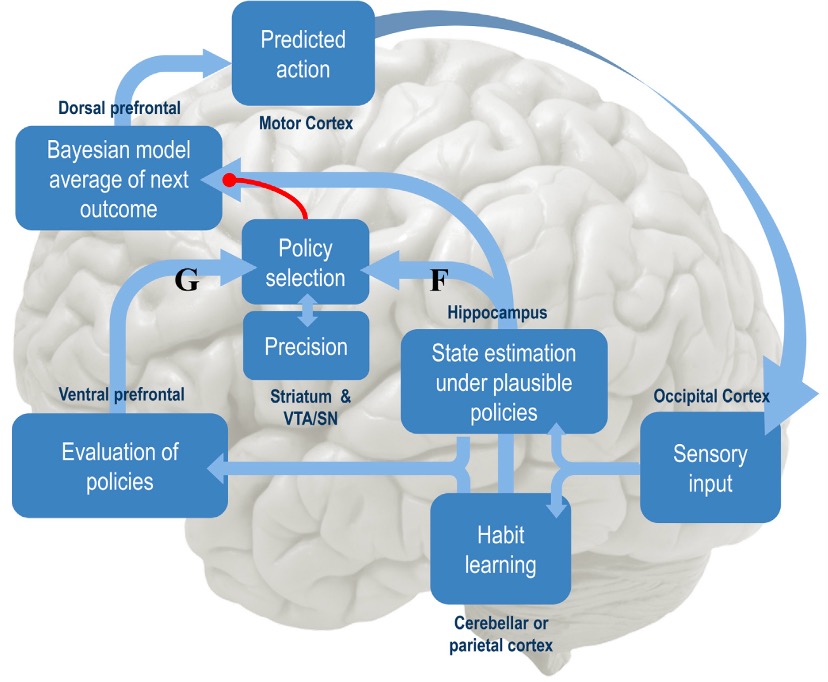}
\caption{Functional brain anatomy of belief updating in active inference: Action-perception loop (large arrow outside the brain) and biologically plausible message passing (arrows inside the brain)~\cite{FFRS16}.}
\label{fig5}
\end{figure}
Figure~\ref{fig5} provides an overview of active inference via the functional brain anatomy implicit in belief updating, highlighting the action-perception loop (large arrow outside the brain, rendering outcomes dependent upon actions) and biologically plausible (i.e., neuroscientifically measurable) message passing between different functional areas inside the brain. As shown in the figure, all the heavy lifting in active inference is done by minimizing free energy, which comes in two flavors: \emph{Variational free energy $F$} and \emph{expected free energy $G$}, as explained next.

\subsection{Free Energy Minimization via Self-Evidencing}
The above described Bayesian belief update in active inference is where \emph{learning} emerges. More precisely, the underlying generative model enables active inference agents to learn the \emph{causal} structures of their worlds instead of merely correlations, endowing them with a so-called \emph{world model}, a capability largely absent in alternative AI/ML algorithms (e.g., reinforcement learning). To see this, it's crucial to better understand the role of variational and expected free energy in active inference.

Recall from above that exact Bayesian inference is often computationally intractable in practical applications of active inference, calling for approximate Bayesian inference instead. Free energy minimization converts an intractable computation problem into an optimization problem that can be solved in a computationally efficient manner. We will show how variational free energy $F$ can be used to solve the POMDP within a given generative model and how expected free energy $G$ extends this approach to infer optimal choices. Simply put, $F$ is a measure of the free energy of the present (implicitly the past thanks to the Markovian property of Markov blankets), while $G$ is a measure of the free energy of the future. Put differently, $F$ and $G$ provide the retrospective view (``how good has this action plan turned out so far?'') and prospective view (``how good do I expect things to go if I continue to follow this action plan'') on pursuing a certain \emph{policy} $\pi$ (i.e., sequence of actions), respectively. 

More formally, an active inference agent aims at finding beliefs in its generative model for which observations provide the most evidence, a process known as \emph{self-evidencing}. Model evidence can be expressed as the sum of the probabilities of observations for every combination of states and policies: $p(o)=\sum_{s,\pi}p(o,s,\pi)$. Next, one can multiply and divide the joint distribution $p(o,s,\pi)$ by an initially arbitrary distribution over states and policies $q(s,\pi)$ that is iteratively updated to match the true posterior distribution $p(s,\pi|o)$ as closely as possible. For mathematical convenience, one can take the negative logarithm of the resulting term, leading to
\begin{equation}
-\text{ln}\ p(o)=-ln\ \sum_{s,\pi}\frac{p(o,s,\pi)q(s,\pi)}{q(s,\pi)}=-\text{ln}\ \mathbb{E}_{q(s,\pi)}\left[\frac{p(o,s,\pi)}{q(s,\pi)}\right],
\end{equation}
where $\mathbb{E}_{q(s,\pi)}[x]$ denotes the expected value or expectation of a distribution $x$ with each value weighted by $q(s,\pi)$. The term $-\text{ln}\ p(o)$, i.e., negative log model evidence, is called surprisal in information theory, or often \emph{surprise} for short. Using Jensen's inequality, which states that the expectation of a logarithm is always less than or equal to the logarithm of an expectation, we arrive at the following definition of variational free energy $F$:
\begin{eqnarray}
-\text{ln}\ p(o) &=& -\text{ln}\ \mathbb{E}_{q(s,\pi)}\left[\frac{p(o,s,\pi)}{q(s,\pi)}\right]\\
&\leq& -\mathbb{E}_{q(s,\pi)}\left[ \text{ln}\ \frac{p(o,s,\pi)}{q(s,\pi)} \right]\\
&=& \mathbb{E}_{q(s,\pi)}\left[ \text{ln}\ \frac{q(s,\pi)}{p(o,s,\pi)} \right]\\
&=&D_{KL}[q(s,\pi)\ ||\ p(o,s,\pi)]=F,
\end{eqnarray}
where $D_{KL}$ denotes the Kullback-Leibler (KL) divergence, also sometimes called \emph{relative entropy}, between the two distributions $q(s,\pi)$ and $p(o,s,\pi)$. $D_{KL}$ is a non-negative measure of their dissimilarity. It is equal to zero if $q(s,\pi)=p(o,s,\pi)$, and greater than zero otherwise. As a consequence, $F$ places an upper bound on surprise. The approximate distribution $q(s,\pi)$ that minimizes $F$ is the one that best approximates the true posterior distribution $p(o,s,\pi)$. Hence, \emph{free energy minimization} helps an active inference agent to maximize model evidence (i.e., $D_{KL}=0$) and minimize surprise of its observations, given that evidence is the inverse of surprise.

To see this, a common way to express variational free energy as placing an upper bound on surprise for a given policy $\pi$ uses the product rule of probability $p(o,s|\pi)=p(s|\pi)p(o|s,\pi)$ to obtain:
\begin{eqnarray}
F(\pi)&=&\mathbb{E}_{q(s|\pi)}\left[\text{ln}\ \frac{q(s|\pi)}{p(o,s|\pi)} \right]\\
&=&\mathbb{E}_{q(s|\pi)}\left[\text{ln}\ q(s|\pi)-\text{ln}\ p(o,s|\pi)\right]\\
&=&\mathbb{E}_{q(s|\pi)}\left[\text{ln}\ q(s|\pi)-\text{ln}\ p(s|\pi)\right]-\nonumber\\
&&\mathbb{E}_{q(s|\pi)}\left[\text{ln}\ p(o|s,\pi) \right]\\
&=&\underbrace{D_{KL}[q(s|\pi)\ ||\ p(s|\pi)]}_\text{Divergence}-\underbrace{\text{ln}\ p(o|\pi)}_\text{Evidence}\\
&=&\underbrace{D_{KL}[q(s|\pi)\ ||\ p(s|\pi)]}_\text{Divergence}+\underbrace{[-\text{ln}\ p(o|\pi)]}_\text{Surprise}
\end{eqnarray}

For decision-making, expected free energy $G$ also needs to be calculated relative to preferences for some sequences of future states and observations that have not yet occurred for each possible policy $\pi$. In active inference, this is formally accomplished by equipping the generative model not only with prior beliefs about states but also with prior expectations over observations, $p(o|\bold{C})$ (see above discussion of $\bold{C}$ in POMDP). These prior expectations over observations play the role of \emph{preferences} $\bold{C}$ definitive of a living organism's phenotype, e.g., seeking warmth when cold, or water when thirsty. Note that this is a central move within active inference: An active inference agent seeks to find policies that are expected to produce those preferred observations $\bold{C}$. To score each possible $\pi$ in this way, expected free energy can be expressed as follows:
\begin{equation}
G(\pi)=\underbrace{D_{KL}[q(o|\pi)\ ||\ p(o|\bold{C})]}_\text{Divergence}+\underbrace{\mathbb{E}_{q(s|\pi)}[H[p(o|s)]]}_\text{Expected Entropy},
\label{expected_entropy}
\end{equation}
where $H[p(o|s)]=-\sum_{o|s}p(o|s)\text{ln}\ p(o|s)=-\mathbb{E}_{p(o|s)}[\text{ln}\ p(o|s)]$ denotes the \emph{entropy} of distribution $p(o|s)$. Note that the term expected entropy in Eq. (\ref{expected_entropy}) denotes the average surprise of future observations for a given policy $\pi$.

\section{Advancing AI: Moving Beyond Brains}
\label{sec:BeyondBrains}
In this section, we elaborate on those mycorrhizal networks that Friston has mentioned as a prime example for developing more advanced AI systems by moving beyond brains, as outlined in his AI vision and R\&D roadmap in Section~\ref{sec:vision}.C. We pay particular attention to the mutually beneficial symbiosis between fungi and flora, since both are species without a brain and thus pose a challenge to translating the theory of FEP into productive scientific practice, as indicated in Section~\ref{sec:vision}.B.

\subsection{Mycorrhizal Networks}
Forests all over the world regulate their natural ecosystems via complex, symbiotic mycorrhizal networks that communicate, nourish, and sustain vast ecosystems. They represent nature's underground Internet popularized under the moniker the \emph{wood-wide web}~\cite{Read97,Whitfield07}. The wood-wide web is formed through mycorrhizal networks by connecting the roots of plants -- belonging to the same or different species -- via underground fungal networks. Mycorrhizal networks are able to mediate plant-plant interactions. Plants acquire carbon from these underground fungal networks by means of exchange of resources between different plants. However, it is important to note that recent outlets in popular media (e.g., high-profile books, newspapers, magazines, documentaries, films, TED talks, podcasts, and even television series) about mycorrhizal networks in forests are not based on scientific evidence and therefore don't help further our understanding of important characteristics of the \emph{structure and function} of mycorrhizal networks (to be further investigated below in Section~\ref{sec:6GWorldBrain}). Even though the transfer of carbon through underground pathways is in line with Simard's seminal paper on the wood-wide web~\cite{SPJM97}, the role of underground pathways other than mycorrhizal networks is often disregarded. Nonetheless, there remain plenty of research opportunities to experimentally investigate the structure and function of mycorrhizal networks in forests. Among others, a deeper understanding of which role the topology of mycorrhizal networks plays in the growth of trees and the resilience of underground fungal networks is needed~\cite{KaJH23}.

The wood-wide web is a rather problematic term since it equates organisms with machines. As a result, we misinterpret organisms as such. Unlike the Internet, the wood-wide web is made up of fungal links that act like \emph{active} fiber-optic cables, i.e., fungi have a life of their own. The fact that fungi are active participants makes a huge difference, especially in the light of active inference. It positions fungi as brokers of entanglement that are able to mediate symbiotic interactions between flora and fungi. The number of fungi species is estimated to be six to ten times larger than flora species in the world. In fact, fungi are the largest living organisms on planet Earth, whereby some are estimated to be as old as 2,400 years. And, importantly, they exhibit the following salient characteristics:
\begin{itemize}
\item{\bf Growth of Mycelium -- Hyphae \& Homing:} \emph{Mycelium} is the most common habit of fungi and is better thought of as a process rather than thing. It may be best viewed as an exploratory, irregular tendency. The vast majority of fungi create networks consisting of cells referred to as \emph{hyphae}, which are only a single cell thick (similar to thin strands of optical fiber). Hyphae are thin structures that tangle into mycelium. The growth of mycelium involves the following two steps: ($i$) First, hyphae branch, and then ($ii$) they fuse. Without fusion, hyphae would be unable to form complex networks. Prior to fusing, hyphae have to find each other by attracting one another. This process is called \emph{homing}, which might be viewed as a process of reducing disorder (i.e., reducing entropy) by feeding upon negative entropy.
\item{\bf Transfer of Resources -- From Areas of Abundance to Areas of Scarcity:} Fungi are decentralized organisms. As a result, coordination in the mycelium happens everywhere at once and nowhere in particular. In many fungal networks, the \emph{transfer of resources is done downhill}, i.e., from areas of abundance to areas of scarcity. In doing so, they create an equilibrium (characterized by increased entropy). Furthermore, larger plants transfer resources to smaller plants. Note that this behavior presents a puzzle. It's not obvious why plants would give resources to a fungus, which in turn gives them to another plant, which may become a potential competitor. 
\item{\bf Mushrooms \& Spores -- Entreating the Visible More-Than-Fungal World:} Hyphae make mycelium, but they also make \emph{mushrooms}, which are fungi's fruiting bodies. They are also the place where fungi produce \emph{spores}, which they use to disperse themselves. Fungi use mushrooms to access the \emph{more-than-fungal world} above ground, thereby rendering fungi visible (by acting as metamorphic agents). Hence, mushrooms enable the embodiment of fungi to help them engage in a certain agency, i.e., sporulating, in the real world (to be revisited in the context of cyberfungi as embodied, enactive AI below in Section~\ref{sec:6GWorldBrain}).
\end{itemize}

It is interesting to note that nature's underground pathways bear some resemblance to the brain's subcortex, ``the subway of the brain'' in Fig.~\ref{fig2}. Similar to myelin, mycelium interconnects different forest (rather than brain) regions and also has the appearance of white matter.

\subsection{The ``Root-Brain'' Hypothesis}
The structure of fungal networks in forests resembles that of the neural networks in human brains. Like neurotransmitters are sent in neural networks, carbon is transmitted in fungal networks between trees. From both neural and fungal networks emerge communication, connection, cohesion, and humans/trees use them to perceive their environment. These similarities have given rise to the so-called \emph{``root-brain'' hypothesis}.

The ``root-brain'' hypothesis was remarked for the first time by Charles Darwin and his son Francis in their revolutionary book \emph{The Power of Movements in Plants}, which departed from the classical and still dominant view of plants as organisms which had no need of movements that were based on sensory perceptions or a brain-like organ -- a view which traces back to Aristotle's concept of placing plants outside the realm of cognitive, animated, animal living systems. The Darwins' ``root-brain'' hypothesis has been forgotten in the literature until 2005, when it was discussed at the first symposium on plant neurobiology held in Florence, Italy. It has been slowly penetrating mainstream research since then, as recent advances in plant molecular biology unmasked plants as sensory and communicative organisms characterized by active, problem-solving behavior. The ``root-brain'' hypothesis claims that the brain-like root apices monitor and integrate numerous parameters simultaneously and then translate these sensory experiences into complex motoric responses, e.g., crawling-like searching movements. This animal-like sensory-motoric circuit allows adaptive behavior (similar to the action-perception loop in active inference). Simply put, the ``root-brain'' hypothesis states that plants are anchored in the soil by their brains, while exposing their sexual organs to the air and to prospective pollinators, e.g., bees. More importantly, it fills a gap in the all-embracing \emph{Living Systems Theory} of James Grier Miller and is also in accord with Sir Jagadish Chandra Bose's \emph{Unity of Life}~\cite{BMVB09}.

In the following section, we are going to explore which role the ``root-brain'' hypothesis may play in realizing the vision of a 6G world brain by advancing AI beyond the human brain.

\section{Toward the 6G World Brain: From Homo Technicus to Interbeing}
\label{sec:6GWorldBrain}
We have seen that active inference is a normative framework to characterize Bayes-optimal behavior. However, it is important to note that the design of the right generative model depends on the desired behavioral outcomes. Each generative model should be associated with different sorts of behavior. As such, more complex behaviors associated with a suitable agency may be accounted for by specifying different generative models that are capable of modelling \emph{alternative futures}, or different ways in which events might play out in the future, and then selecting among them~\cite{PaPF22}.

In the remainder of this paper, we are going to develop a generative model that allows us to knit our own \emph{socio-technological cocoon} for the human-AI co-evolution of homo technicus to become a metamorphic active inference agent of the future 6G world brain and global mind.

\subsection{Homo Technicus \& Interbeing: The Human Strategy}
In the new Age of AI, briefly mentioned in Section~\ref{sec:intro}, the human-AI co-evolution of \emph{homo technicus} will lead to a human species that may live in symbiosis with machine technology~\cite{KiMS24}. In conceptually unlocking the possibility of a cyborganic entanglement of these two species -- one synthetic, the other organic -- reveals the need for creating a world in which AI becomes more like us, or one in which we become more like AI. In its evolutionary speed and diversification, AI's development will be akin to the Cambrian explosion: the emergence of a wide variety of hybrid cyborganic lifeforms within a single, highly compressed period of time relative to the preceding epoch.

To help homo technicus cope with emerging advanced AIs, he or she ought to apply \emph{the human strategy} laid out by MIT's Alex Pentland~\cite{Pentland14}. Pentland suggests that today's AI is as far from Norbet Wiener's original notion of cybernetics as you can get, because it is not contextualized; it is -- what he calls -- a little idiot savant. Conversely, humans can perceive things in a broadly competent way and thus have a commonsense understanding of the real world that they bring to most problems. \emph{So what would happen if we replaced AI neurons with people?} Pentland calls this the human strategy. According to Pentland, on the horizon is a next-generation AI vision of how we can make humanity more intelligent by building a human AI, where social interactions such as  social learning (i.e., spreading ideas and transforming those ideas into behaviors) are the primary forces driving the evolution of collective intelligence. This conclusion is echoed by Max Borders through his concept of the \emph{human hive mind}~\cite{Borders18}. According to Borders, more and more, humans act as neurons with decentralized blockchain technologies acting as connective tissue to create programmable incentives, which are technically known as blockchain \emph{tokens}. 

\begin{figure*}[t]
\centering
\includegraphics[width=\linewidth]{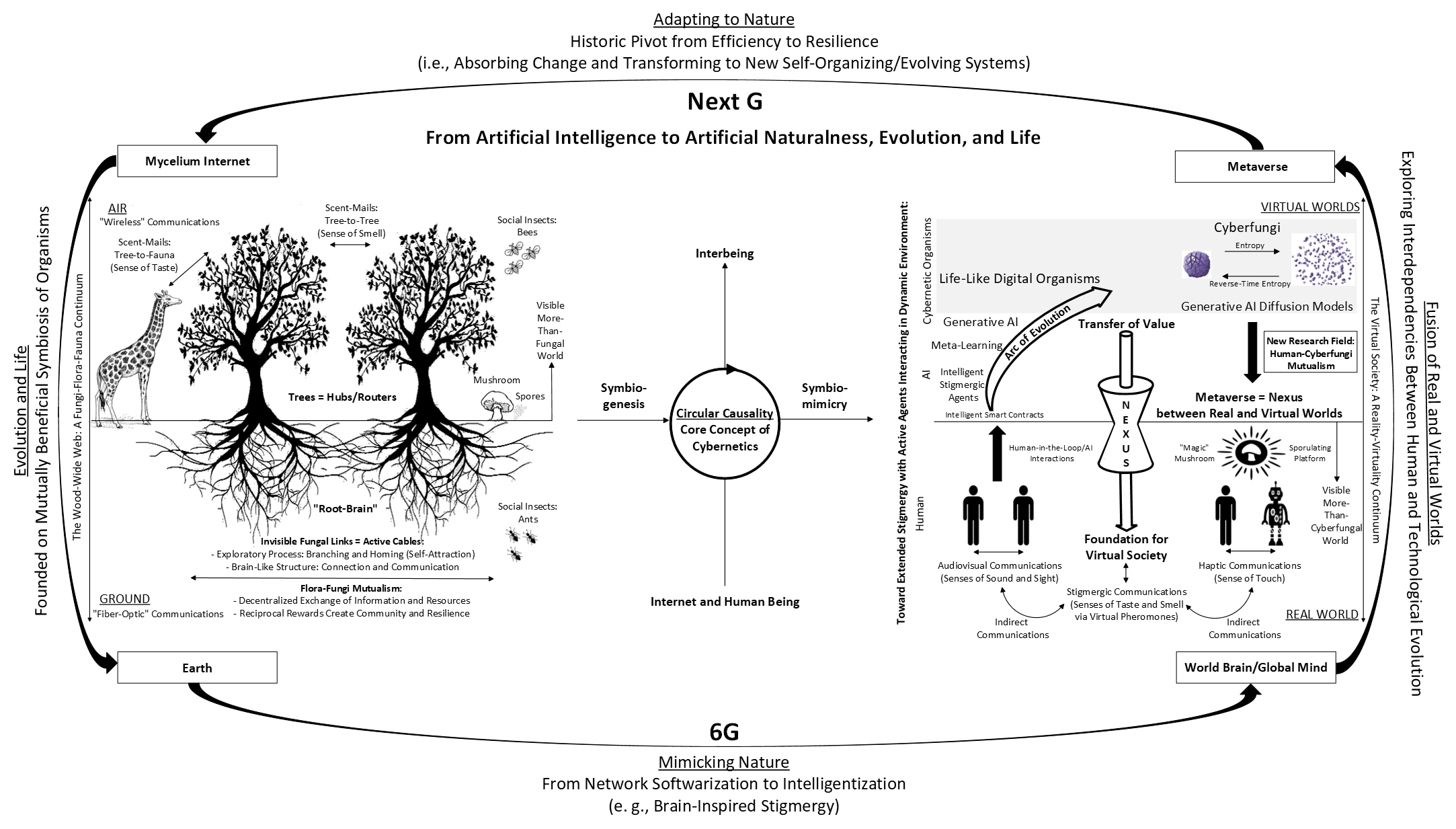}
\caption{6G transformation of Earth into World Brain/Global Mind by mimicking nature (i.e., biomimicry) via human-AI interaction of natural and artificial intelligence networked into a cyber-physical ecosystem of real and virtual worlds: Forests regulate their natural ecosystems via underground mycorrhizal fungal networks for the exchange of resources between trees. \emph{Circular causality} -- the core concept of cybernetics (see center of figure) -- is used to biomimic the ``root-brain'' and metamorphic fungi-turned-mushroom for the mutually beneficial symbiosis between Inter(net) and (human) being, giving rise to the powerful concept of \emph{Interbeing}~\cite{MaHS24}.}
\label{fig6}
\end{figure*}
Recently, in~\cite{MaHS24}, we have elaborated on the importance of a rich techno-social environment that triggers shared, collective experiences of the human hive mind, where a group has the feeling that they are moving together toward the same or complementary goals. We have put a particular focus on the unifying design of virtual, embodied, intelligent cross-reality environments based on the mutually beneficial symbiosis between Inter(net) and (human) being, giving rise to the powerful concept of \emph{Interbeing}.\footnote{{\emph{Interbeing} is a word that was originally coined by the Buddhist monk Thich Nhat Hanh to describe the mutual interdependence in human relationship and humanity's relationship to the natural world---a connection to what many refer to as the life force, a higher power or purpose. In \emph{The Art of Living}, Thich Nhat Hanh states that the word is not in the dictionary yet.}} Interbeing enables active agents such as homo technicus to interact in a dynamic human-AI-cybernetic organisms environment. In the following, we are going to demonstrate how the concept of Interbeing may be leveraged for realizing the dyadic interaction between human and cyberfungi active inference agents introduced above in Fig.~\ref{fig3}, while adhering to the human strategy to foster the human-AI co-evolution of homo technicus, tying the real and virtual worlds of the future phygital environments such as the 3D Metaverse tighter together, all in an attempt in incentivizing desired behavioral outcomes and alternative futures via purpose-driven tokens (purpose free to be chosen by active inference agents, see below for example of social learning via spreading ideas and actuating tokens by transforming those ideas into desirable new social norms).

\subsection{Symbiogenesis \& Circular Causality: Teleology}
Recall from Section~\ref{sec:intro} that in the face of the growing complexity of today's communication networks we have to completely change them from being static into being a living entity equipped with a network `brain' by applying 6G's mainstay approach of Net4AI (i.e., network for AI). Figure~\ref{fig6} illustrates our envisioned transformation of Earth into the 6G world brain and global mind by mimicking nature's more-than-human intelligence (see arrow at bottom of figure). The left-hand side of Fig.~\ref{fig6} depicts the ``root-brain'' as well as the metamorphic fungi-turned-mushroom, entreating the more-than-fungal world above ground to disperse themselves via spores. Flora-fungi mutualism, i.e., the mutually beneficial symbiosis between flora and fungi, exemplifies the evolutionary process of \emph{symbiogenesis} which merges two separate organisms to form a single new organism. Biomimicking the evolution of new, more complex integrated organisms via symbiosis of Internet and human being (Interbeing) toward a single, higher-level organism is particularly promising, an approach known as \emph{symbiomimicry}. 

Here, we are interested in forming the new cyborganic entity of homo technicus and their human-AI co-evolution via our proposed human-cyberfungi mutualism premised on active inference. According to~\cite{PaPF22}, active inference is closely related to cybernetic ideas about the purposeful, goal-directed nature of behavior and the importance of feedback-based agent-environment interactions. Specifically, we are interested in applying \emph{circular causality} as the fundamental principle for the control and management of complex adaptive systems such as the emerging Metaverse (see center of Fig.~\ref{fig6}). Norbert Wiener introduced circular causality as the core concept of cybernetics, which he used as a generalized feedback mechanism for feedback-controlled purpose, referred to as \emph{teleology} (from Greek \emph{t\'elos}: Purpose, goal, or ultimate end). Hence, it denotes a purpose-driven (teleological) mechanism. Circular causality takes the observed outcomes of actions as feedbacks for steering subsequent actions in support of maintaining particular conditions such as homeostasis of systems. It is worth mentioning that Wiener introduced circular causality as a teleological mechanism for regulatory and purposed systems in both machines and living organisms.

\subsection{6G World Brain: Metamorphic Generative Model for Knitting Our Own Socio-Technological Cocoon}
As shown on the right-hand side of Fig.~\ref{fig6}, in future 6G networks humans may continue to engage in conventional human-to-human audiovisual communications, involving the senses of sound and sight. They may also remotely steer robots thru haptic human-to-robot communications involving the sense of touch, leading to embodied communications. Both human-to-human and human-to-robot interactions are illustrative examples of direct communications. In addition, humans may partake in indirect communications mediated through stigmergic communications by mimicking the senses of taste and smell via virtual pheromones in the online environment. For further technical details on haptic and stigmergic communications in 6G, we refer the interested reader to our JOCN special issue OFC 2021 paper~\cite{MEBR22}.

Building on our previous work on embodied (human-to-robot) interaction, the focus of this paper lies on demonstrating more advanced human-AI interactions networked into a cyber-physical ecosystem of artificial and natural intelligence, in which humans are integral participants, as envisioned by Friston \emph{et al.} in Section~\ref{sec:vision}. Human-in-the-loop/AI interactions let humans interact with different types of AI, ranging from intelligent smart contracts operating on simple if-then programming to more advanced \emph{generative AI}, which in turn will pave the way to future life-like digital organisms, as indicated by the arc of evolution in Fig.~\ref{fig6} (see right-hand side of figure). 

Here, we are interested in adapting and extending generative AI to help us implement our proposed metamorphic cyberfungi active inference agent, briefly introduced in Fig.~\ref{fig3}, as an early example of future cybernetic organisms. Specifically, we adapt and extend \emph{OpenAI's improved Denoising Diffusion Probabilistic Model (DDPM)}\footnote{OpenAI improved Denoising Diffusion Probabilistic Model (DDPM) codebase available at: \texttt{https://github.com/openai/improved-diffusion}} to biomimic the salient features of mycorrhizal networks, paying particular attention to their two important characteristics of structure and function, as explained in Section~\ref{sec:BeyondBrains}.A. Recall that in active inference, the generative model is supposed to closely biomimic the underlying generative process found in forests (see also Fig.~\ref{fig4}). To do so, we use the forward diffusion process of DDPM to biomimic the branching process and exploratory, irregular tendency of hyphae, while the backward denoising process of DDPM is used to to biomimic the fusing and homing process of hyphae to create complex mycelium-like networks. Note that the use of generative AI diffusion models such as DDPM allows us to create not only an increase in entropy (diffusion) but also decrease or reverse-time entropy (denoising), which may be viewed as negative entropy. Recall from Section~\ref{sec:AIF} that entropy denotes the average surprise, which needs to be minimized in active inference. Importantly, we extend OpenAI's improved DDPM -- which in its original form is a completely disembodied AI -- to let our cyberfungi entreat the physical more-than-cyberfungal world in order to dissiminate cybernetic spores (i.e., teleological tokens), similar to their metamorphic fungi-turned-mushroom counterparts (see right-hand side of Fig.~\ref{fig6}). As a result, DDPM becomes an embodied AI as well as enactive AI, which are two main characteristics of active inference agents (see Section~\ref{sec:AIF}.A).

\begin{figure}[t]
\centering
\includegraphics[width=1.05\linewidth]{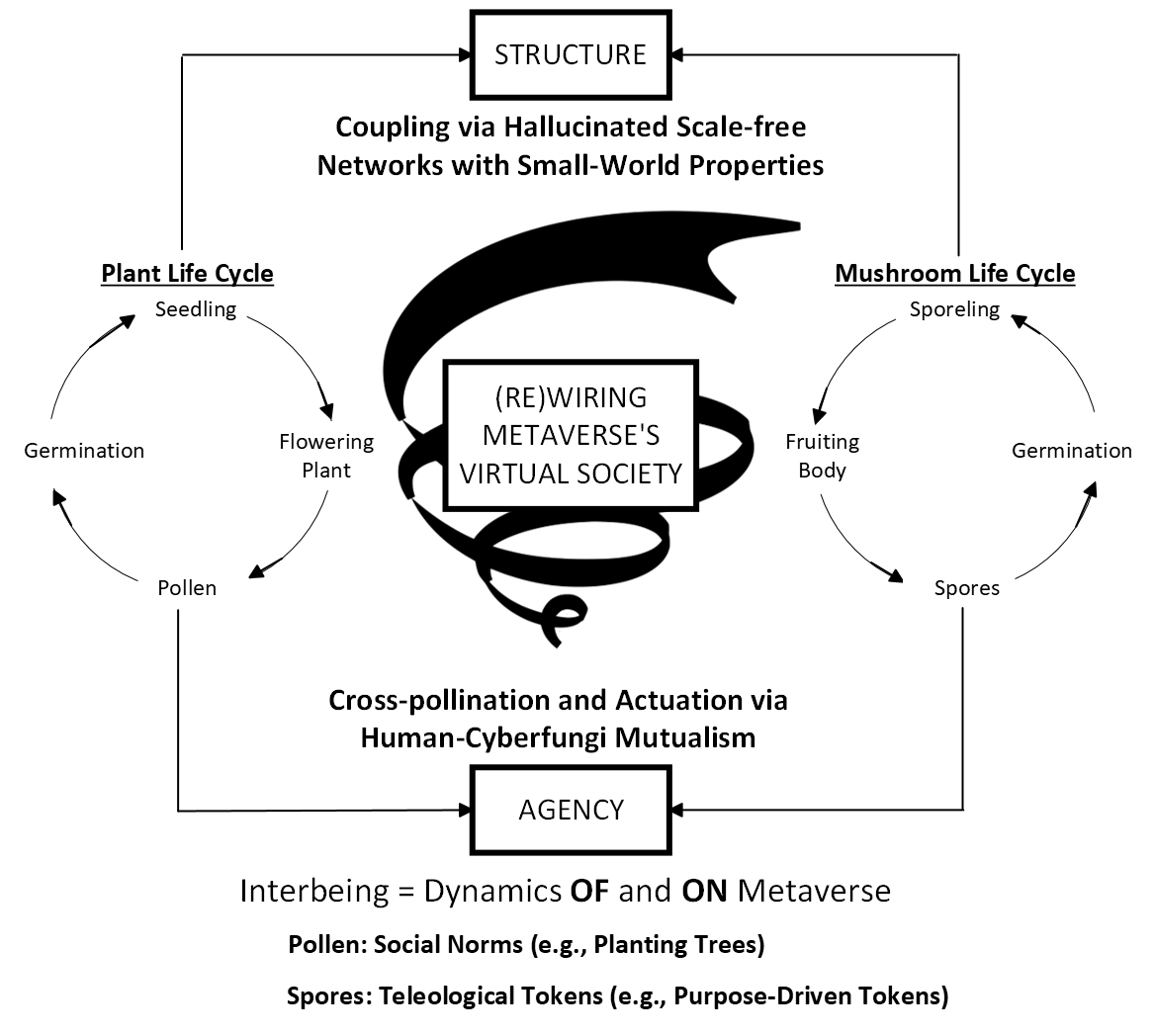}
\caption{Schematic illustration of proposed metamorphic generative model for cyberfungi active inference agent symbiomimicking mediated interactions between plant and mushroom life cycles for enactive, embodied AI: ($i$) \emph{Structure}: Generative model dynamically couples pairs of humans via hallucinated scale-free networks with small-world properties that are widely found in biological, social, and man-made systems; and ($ii$) \emph{Agency}: Similar to cross-pollination among plants, pairs of directly linked humans spread a desirable new social norm incentivized by reciprocal rewards in the form of teleological (purpose-driven) tokens, which are dispersed into the physical more-than-cyberfungal world.}
\label{fig7}
\end{figure}
Figure~\ref{fig7} schematically illustrates our proposed metamorphic generative model for cyberfungi, putting a particular focus on explaining their underlying network \emph{structure} and the resulting networked \emph{agency} (i.e., function), two important characteristics of mycorrhizal networks in need for further understanding (see Section~\ref{sec:BeyondBrains}.A). Towards this end, we exploit DDPM's well-known capability of \emph{hallucination}.\footnote{\emph{Hallucinate} is Cambridge Dictionary's Word of the Year 2023, following a year-long surge in interest in generative AI with public attention shifting towards its limitations and whether they can be overcome through `grounding' by means of human feedback. Clearly, the importance of human `grounding' calls for human-in-the-loop/AI interactions.} More specifically, we let our cyberfungi generative model hallucinate so-called scale-free networks with small-world properties to couple the biomimicked life cycles of plants and mushrooms. Now, it's important to note that scale-free networks can be widely found in nature, including our human brain. The nodal degree distribution of scale-free networks follows a power law, whereby a few hubs have many connections that provide these networks with the so-called small-world property, which is a quantifiable characteristic widespread in biological (e.g., neuronal networks), social (e.g., social networks), and man-made systems (e.g., world wide web). The resultant small-world, scale-free networks have an increased wiring efficiency and are therefore biologically more economical than random networks, often with important dynamical consequences such as different forms of biological contagion (and social contagion, as we shall see shortly). The study of scale-free networks with small-world property also plays an instrumental role in the \emph{emergence of spontaneous order}, i.e., the transition from randomness to order, in self-organizing organisms widely found in nature~\cite{Strogatz03}. In our comprehensive search of over 7,000 network datasets, we haven't found a single dataset for small-world, scale-free networks. Hence, we used NetworkX to generate a dataset consisting of 1,304,483 unique tensors, each defining a different small-world, scale-free synthetic network, for training OpenAI's improved DDPM.\footnote{NetworkX is a Python package that lets users create, manipulate, and study the structure, functions, and dynamics of complex networks, whose nodes and edges can be attributed with arbitrary data (e.g., weights or time series). For further details, please visit: \texttt{https://networkx.org}.}

As for the agency emerging from these underlying scale-free networks with small-world properties, the generative model lets human beings and cyberfungi -- both metamorphic agents (see Section~\ref{sec:AIF}.A) -- engage in novel types of human-AI interaction to knit our own \emph{socio-technological cocoon} for the human-AI co-evolution of the 6G world brain, where humans and cyberfungi act as neurons networked into a cyber-physical ecosystem of artificial and natural intelligence to form the new cyborganic entity homo technicus premised on human-cyberfungi mutualism, a new research field as shown in Fig.~\ref{fig6} (see right-hand side of figure). Circular causality is used to take the observed outcomes of human behavior as feedbacks for steering subsequent cyberfungi actions in support of driving the evolution of collective intelligence by incentivizing social learning, i.e., spreading ideas and transforming those ideas into desirable new social norms.

Toward this end, our proposed metamorphic generative model facilitates human-cyberfungi mutualism by biomimicking the agency of the plant and mushroom life cycles. While the plant life cycle produces pollen for cross-pollination between plants, the mushroom life cycle disperses spores that need to be \emph{actuated} in order to germinate (see Fig.~\ref{fig7}). To do so, humans interact with cyberfungi by connecting to their created small-world, scale-free networks such that one human performing a desirable new social norm (e.g., planting trees) cross-pollinates other connected humans such that the new social norm starts spreading from human to human with a certain spreading probability denoted by $0\leq r\leq 1$ (a social innovation phenomenon known as social contagion). Each human who starts adapting the new social norm is rewarded with cybernetic spores, which get actuated by the human's action. These cybernetic spores can be any type of teleological tokens such as purpose-driven tokens widely used in decentralized blockchain networks. This human-cyberfungi mutualism consisting of the cyclic cross-pollination of social norms and actuation of teleological tokens (cybernetic spores) continuously repeats itself. In doing so, it (re)wires the emerging 6G world brain comprising interacting human and cyberfungi active inference agents. Note that social norms act as entropy-reducing devices for humans pursuing a desirable purpose by reducing behavioral chaos, as intended by active inference as normative framework to characterize Bayes-optimal behavior. 

\begin{figure}[t]
\centering
\includegraphics[width=\linewidth]{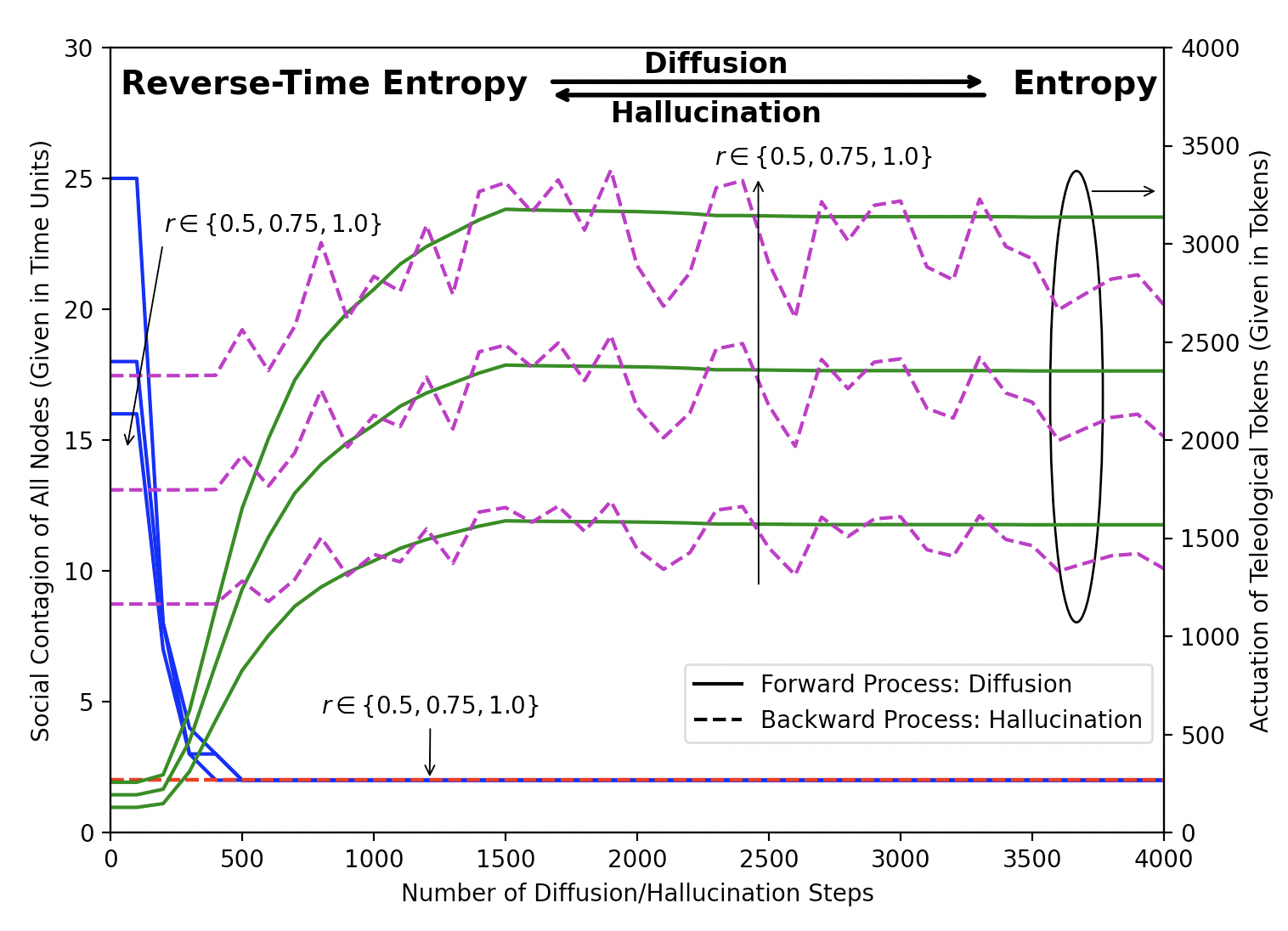}
\caption{Impact of cyberfungi generative model on entropy, which denotes the average surprise that needs to be minimized in active inference: Adopted DDPM allows to create not only an increase in entropy (diffusion) but also decrease or reverse-time entropy (denoising), which may be viewed as \emph{negative entropy} living organisms feed upon (see Erwin Schr\"odinger in Section~\ref{sec:vision}.B). The forward diffusion process of DDPM is adopted to biomimic the branching process and exploratory, irregular tendency of hyphae, while the backward denoising process of DDPM is adopted to to biomimic the fusing and homing process of hyphae for the hallucination of complex mycelium-like network structures.}
\label{fig8}
\end{figure}
Figure~\ref{fig8} demonstrates the beneficial role our cyberfungi play in acting as brokers of cyborganic entanglement and mediating mutually beneficial human-AI interactions. In general, an information cascade is needed to create all sorts of collective behavior outbreak, whereby small-world networks are particularly suited to foster social contagion. The x-axis of Fig.~\ref{fig8} depicts the number of diffusion steps of the forward process (i.e., from left to right) and the number of hallucination steps of the backward process (i.e., from right to left), respectively. For illustration, let us consider a regular 64-node ring lattice that the adopted DDPM diffuses into a random network by increasing its entropy (disorder). We assume that initially only one node, connecting a single human or a cluster of multiple humans to the network, adapts a desirable new social norm (e.g., planting trees). A node that is directly linked to that initial node copies the new social norm with spreading probability $r\in\{0.5, 0.75, 1.0\}$during each time unit. The left-hand y-axis denotes the required number of time units it takes until the new social norm propagates throughout the entire population of all 64 nodes as a function of diffusion/hallucination steps. Figure~\ref{fig8} depicts that the initial regular 64-node ring lattice (without any diffusion/hallucination) requires up to 25 time units, depending on the value of $r$. Clearly, for increasing $r$ the number of required time units decreases, reaching 16 time units for $r=1.0$. The diffusion forward process reduces the number of required time units until it reaches the minimum of 2 after roughly 500 diffusion steps for all values of $r$. Interestingly, in the backward process, the cyberfungi hallucinate scale-free networks with small-world properties such that the number of required time units to achieve social contagion of all nodes remains constantly low at the minimum of 2, independently of the actual value of spreading probability $r$. This is due to the superior wiring efficiency of the hallucinated small-world, scale-free networks favoring social contagion.

While the spreading of a desirable new social norm may be viewed as a value per se, the right-hand y-axis of Fig.~\ref{fig8} denotes the number of actuated teleological tokens (cybernetic spores) each pair of nodes receives from our cyberfungi generative model as a reward for adopting the new social norm. That is, each pair of directly linked nodes that successfully spread the social norm for a given value of $r$ is reciprocally rewarded with a couple of teleological tokens, one for each node. Figure~\ref{fig8} shows that the forward diffusion process increases the number of actuated teleological tokens until it reaches a certain maximum, which differs for varying values of $r$. Note that the initial regular 64-node ring lattice (without any diffusion/hallucination) yields a rather small number of actuated teleological tokens of roughly 300 or less. Conversely, the backward hallucination process is able to achieve a significantly larger number of actuated teleological tokens up to almost 2,500 for $r=1.0$, thus clearly demonstrating a significantly increased number of embodied AI actions entreating the physical more-than-cyberfungal realm.

\section{Conclusions}
\label{sec:conclusions}
To cope with the unprecedented growth in complexity of today's networks, two opposing paradigm shifts have been emerging. One perceives AI/ML techniques as a paradigm shift that allows us to discover hidden relations by inferring, from data obtained by various types of monitors, useful characteristics that could not be easily or directly measured (see excellent overview in~\cite{MRNM19} for details). However, this approach may eventually backfire since AI/ML adds yet another man-made layer to our man-made networks, potentially further increasing their overall complexity, especially in light of unresolved open key AI challenges related to training, learning, and explainability. More importantly, today's \emph{passive AI} is profoundly lacking the key elements found in living intelligent systems. Conversely, the alternative paradigm shift of active inference not only resolves the open key AI challenges but also advocates an \emph{active AI} that views today's complex networks as living organisms rather than man-made machines.

This paper introduced the biomimetic mathematical framework of active inference as a key to true AI and 6G world brain by bridging the gap between artificial and natural intelligence. We showed which role Charles Darwin's ``root-brain'' hypothesis, which has been forgotten in the literature until 2005, may play in advanced human-AI interaction of natural and artificial intelligences networked into a cyber-physical ecosystem premised on active inference. Further, we elaborated on the free-energy principle as a unifying theory of brain, life, and behavior as well as Markov blanketed systems found in nature for the interface design of active inference agents and their metamorphosis, while paying close attention to the central role of entropy in knitting our own socio-technological cocoon for the genesis and evolution of homo technicus, a human species that lives in symbiosis with emerging AIs. We demonstrated how to tackle the central challenge of active inference by developing a generative model that does not emulate the human brain, but describes the problem at hand: Modelling alternative futures for the human-AI co-evolution of homo technicus to become a metamorphic agent of the future 6G world brain, where AI neurons are replaced with humans and cybernetic organisms.

In the new Age of AI, the authors of~\cite{KiMS24} -- including the two eminent technologists Craig Mundie (former chief research and strategy officer of Microsoft) and Eric Schmidt (former CEO and chairman of Google) -- argue that the best path forward for the development of machine intelligence depends on whether AI's structures continue to resemble the structures of the human brain or whether \emph{``the human brain, therefore, becomes not the goal, and not a blueprint, but a midpoint and an inspiration toward something greater.''} The 6G world brain?

\section*{Acknowledgement}The author would like to thank Nika Hosseini for her contribution to the numerical work.

\section*{Funding}Natural Sciences and Engineering Research Council of Canada (NSERC) (Discovery Grant 2021-03224).

\bibliography{sample}

\section*{Author Biography}

\setlength\intextsep{0pt}

\begin{wrapfigure}{L}{0.2\textwidth}
\includegraphics[width=0.23\textwidth]{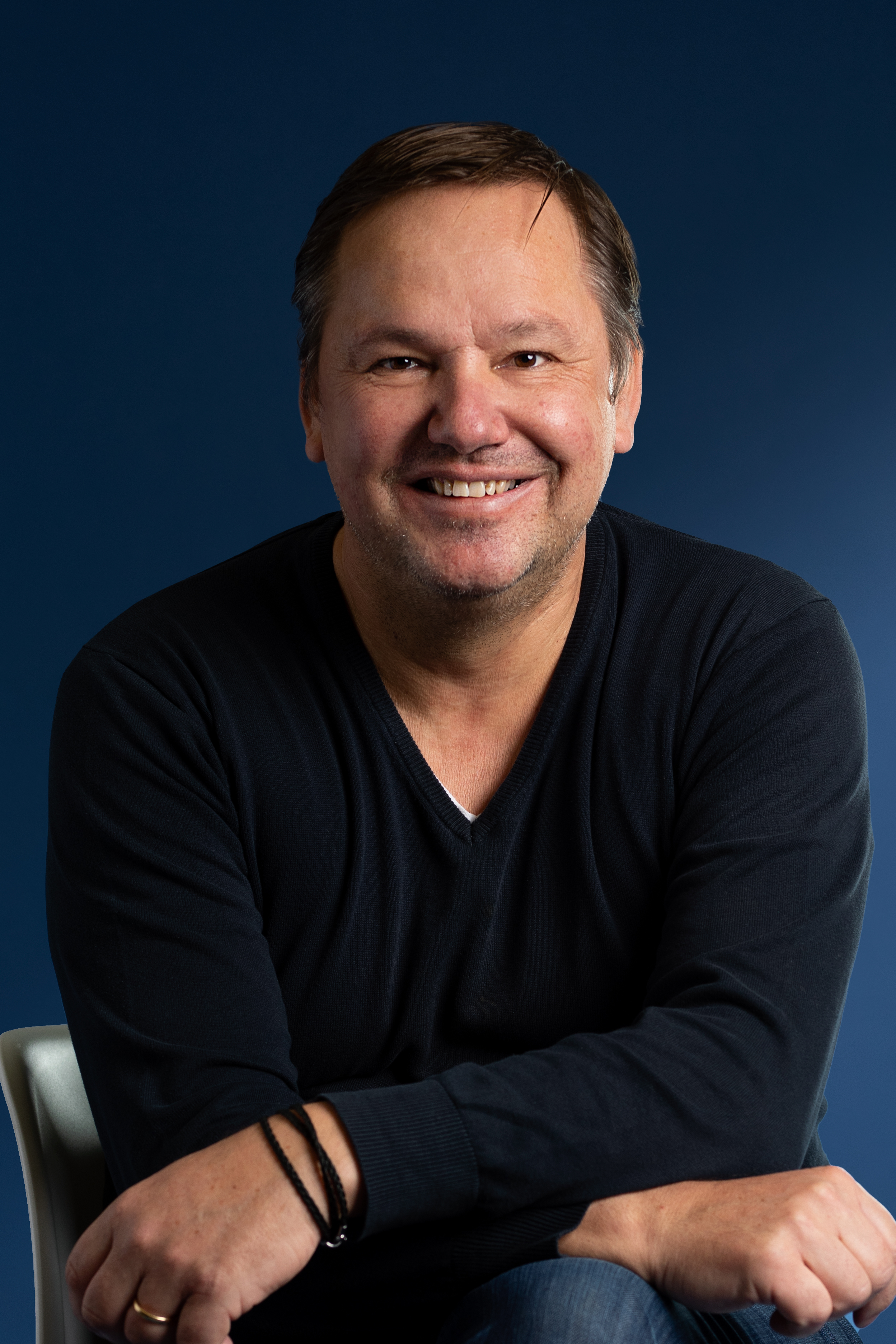}
\end{wrapfigure}
\paragraph{}
\noindent 
\textbf{Martin Maier} is a full professor with INRS, Montr\'eal, Canada. He was educated at the Technical University of Berlin, Germany, and received MSc and PhD degrees both with distinctions (summa cum laude) in 1998 and 2003, respectively. In 2003, he was a postdoc fellow at the Massachusetts Institute of Technology (MIT), Cambridge, MA. He was a visiting professor at Stanford University, Stanford, CA, 2006 through 2007. He was a co-recipient of the 2009 IEEE Communications Society Best Tutorial Paper Award. Further, he was a Marie Curie IIF Fellow of the European Commission from 2014 through 2015. In 2017, he received the Friedrich Wilhelm Bessel Research Award from the Alexander von Humboldt (AvH) Foundation in recognition of his accomplishments in research on FiWi-enhanced mobile networks. In 2017, he was named one of the three most promising scientists in the category ``Contribution to a better society'' of the Marie Sklodowska-Curie Actions (MSCA) 2017 Prize Award of the European Commission. In 2019/2020, he held a UC3M-Banco de Santander Excellence Chair at Universidad Carlos III de Madrid (UC3M), Madrid, Spain. Recently, in December 2023, he was awarded with the 2023 Technical Achievement Award of the IEEE Communications Society (ComSoc) Tactile Internet Technical Committee for his contribution on 6G/Next G and the design of Metaverse concepts and architectures as well as the 2023 Outstanding Paper Award of the IEEE Computer Society Bio-Inspired Computing STC for his contribution on the symbiosis between INTERnet and Human BEING (INTERBEING). Based on Stanford University's list of ``World's Top 2\%'' most cited scientists, he ranks among the top 2\% of all scientists worldwide and has been recently awarded 2024 Highly Ranked Scholar Lifetime status by ScholarGPS as \#2 worldwide in the area of access network (top 0.05\%). He is co-author of the book ``Toward 6G: A New Era of Convergence'' (Wiley-IEEE Press, January 2021) and author of the sequel ``6G and Onward to Next G: The Road to the Multiverse'' (Wiley-IEEE Press, February 2023).\\

\end{document}